\documentstyle[psfig,preprint,eqsecnum,tighten,floats,aps]{revtex}
\makeatletter
\def\d{{\rm d}}
\def\={\mathrel{\widehat\mathalpha{=}}}
\def\be{\begin{equation}}
\def\ee{\end{equation}}
\begin{document}
\preprint{\vbox{\baselineskip=12pt
\rightline{ICN-UNAM-00/15}
\rightline{gr-qc/0011084}
}}

\title{Hair from the Isolated Horizon Perspective}

\author{Alejandro Corichi and Daniel Sudarsky}

\address{Instituto de Ciencias Nucleares, Universidad Nacional Aut\'onoma de 
M\'exico,\\ A. Postal 70-543, M\'exico D.F. 04510, M\'exico.
\\E-mail: corichi@nuclecu.unam.mx, sudarsky@nuclecu.unam.mx}


\maketitle

\begin{abstract}
The recently introduced Isolated Horizons (IH) formalism
has become a powerful tool for realistic black hole physics. In
particular, it generalizes the zeroth and first laws
of black hole mechanics in terms of quasi-local quantities and serves
as a starting 
point for quantum entropy calculations. In this note
we consider theories which admit hair, and analyze  
some new results that the IH provides, when considering
solitons and stationary solutions. Furthermore, the IH formalism allows to
state uniqueness conjectures (i.e. horizon
`no-hair conjectures') for the existence of solutions.
\end{abstract}

\section{Introduction}
The vast majority of analytical work on black holes in general
relativity centers around event horizons in globally stationary
spacetimes\cite{mh}. Even when this is a natural starting
point, it is not entirely satisfactory from a physical viewpoint.
For instance, the collapse to form a black hole, black hole mergers, etc. are
situations not described by stationary solutions. In recent years,
a new framework tailored to consider situations in which the
black hole is in equilibrium (nothing falls in), but which allows for the
exterior region to be dynamical, has been developed. This
{\it Isolated Horizons} (IH) formalism is now in the position of
serving as starting point for several applications, from the
extraction of physical quantities in numerical relativity to
quantum entropy calculations\cite{prl}.
  
The basic idea is to consider space-times with an interior boundary
(to represent the horizon), satisfying quasi-local boundary conditions
ensuring that the horizon remains `isolated'. Although the boundary
conditions are motivated by geometric considerations, they lead to a well 
defined action principle and Hamiltonian framework. 
Furthermore, the boundary
conditions imply that certain `quasi-local charges', defined at the horizon,
remain constant `in time', and can thus be regarded as
the analogous of the global charges defined at infinity in the
asymptotically flat context. The isolated horizons Hamiltonian framework allows
to define the notion of {\it Horizon Mass} $M_\Delta$, as function of
the `horizon charges'. In the Einstein-Maxwell and Einstein-Maxwell-Dilaton
systems considered originally\cite{abf1}, the horizon mass satisfies a
Smarr-type formula and a generalized first law in terms of
quantities defined exclusively at the horizon (i.e. without any reference to
infinity).

The introduction of non-linear matter fields like the Yang-Mills field
brings unexpected subtleties to the formalism\cite{ac:ds}. Firstly,
the Horizon Mass can no longer be written in terms of a Smarr formula
and the first law has to be reconsidered. Second, the
IH formalism seems to be robust enough to allow for new results even
in the static sector of the theory under consideration. The purpose
of this short note is to review these results and to direct to the
relevant literature where details can be found. The structure of this note 
is as follows. In Section~\ref{sec2} we consider the first law.
In Sec.~\ref{sec3} we state the uniqueness and completeness conjectures
in terms of quasi-local charges. In Sec.~\ref{sec4} we discuss a 
formula relating black holes and solitons of the theory. We end with
a summary in Section~\ref{sec5}

\section{The First Law}
\label{sec2}

An isolated horizon is a non-expanding null surface generated by a
(null) vector field $l^a$. The IH boundary conditions imply that
the acceleration $\kappa$ of $l^a$ ($l^a\nabla_al^b=\kappa l^b$)
is constant on the horizon $\Delta$. However, the precise value it takes
on each point of phase space (PS) is not determined a-priori. On the
other hand, it is known that for each vector field $t^a_o$ on spacetime,
the induced vector field $X_{t_{o}}$ on phase space is Hamiltonian
if and only if there exists a function $E_{t_{o}}$ such that
$\delta E_{t_{o}}=\Omega (\delta,X_{t_{o}})$, {\it for any
vector field $\delta$ on PS}. This condition can be re-written
 as\cite{afk},
\be
\delta E_{t_{o}}=\frac{\kappa_{t_{o}}}{8\pi G}\,\delta a_{\Delta}
+ {\rm work\;\; terms}
\ee
Thus, the first law arises as a necessary and sufficient condition for
the consistency of the Hamiltonian formulation. Thus, the allowed vector
fields $t^a$ will be those for which the first law holds.
Note that there 
are as many `first laws' as allowed vector fields $l^a\=t^a$ on the horizon.
However, one would like to have a {\it Physical First Law}, where the
Hamiltonian $E_{t_{o}}$ be identified with the `physical mass' $M_{\Delta}$
of the horizon. This amounts to finding the `right $\kappa$'.
This `normalization problem' can be easily overcome in
the EM system\cite{abf1}. In this case, one chooses the function
$\kappa=\kappa(a_\Delta, Q_\Delta)$ as the function that a {\it static}
solution with charges $(a_\Delta, Q_\Delta)$ has.
However, for the EYM system, this procedure is not as straightforward.
At present, there are two viewpoint towards this issue:
i) If one wants to have a `global normalization' for $l^a$ valid on
all PS, and therefore, a `canonical horizon mass' $M_\Delta$, then
one has to restrict the allowed variations $\tilde{\delta}$ to certain 
directions tangent to some preferred
`leaves' on phase space\cite{ac:ds}. ii) One 
abandons  the notion of
a globally defined horizon mass, but then the first law is valid for
arbitrary variations $\delta$ on PS\cite{afk}. At present
both viewpoints seem to be complementary.

\section{Conjectures}
\label{sec3}
The general prescription for arriving at an
explicit  expression for
$M_{\Delta}$, for general isolated horizons, involves the fixing of the
quantity $\kappa$ as function of the horizon parameters. For this, one
requires some input from the Static solutions.  The first requirement is
that there be no ambiguities. Thus we have to conjecture that ($C1$):
{\it All static BH solutions are characterized by its
horizon parameters arising from the `isolated horizon'
framework}.
In theories where no hair is present, as is the case of the
Einstein-Maxwell-Dilaton system, the number of `quasi-local
charges' equals the number of parameters at infinity labeling
the static solutions\cite{abf1}.
Thus, stating a uniqueness conjecture in this
theory is insensitive as to whether one is postulating it in terms
of quantities at infinity (the standard viewpoint), or in terms
of `quasi-local charges'. Our proposal is that, for general
theories, one should state the  postulate in
terms of purely quasi-local quantities.
In the EYM system, the quasi-local charges are $a_\Delta$, the horizon
area, $Q_\Delta$ and $P_\Delta$, the horizon electric and magnetic charges
respectively. In this case the first conjecture $C1$ reads:
Given a triple of parameters $(a_\Delta,Q_\Delta,P_\Delta)$ for
which a Static solution exists, then the solution is unique. Note that this
provides for a way to formulate uniqueness statements regarding  black
holes that was absent in the theories that admit hairy solutions.

 However, this is not sufficient in  order to have
the Isolated Horizon framework  working for the EYM system to the same
extent that it works,
say, for the Einstein Vacuum, EM, and EM-Dilaton systems.
In order to achieve that, we would need to have a canonical
normalization of $\kappa$
for all the values of the Isolated Horizon parameters. In the
previously mentioned cases this canonical choice is given by
the existence of
static (and spherically symmetric) Black Hole  solutions for all
isolated horizon  values of the parameters.

For the case of the EYM system, in   the regime of staticity
and spherical symmetry there are,
given a fixed value of $a_\Delta$, only a
discrete set of values of $P_\Delta$ for which there are black
hole solutions.
Moreover, within this regime there are
 no  Black Hole solutions for any value of
$P_\Delta \not = 1, 0$ and $Q_\Delta \not =0$. Thus if we want to have
any hope
that the Conjecture might be true we must formulate it outside this
restrictive regime.
Indeed the fact that in EYM systems there are static Black Hole solutions
that are not spherically symmetric,
already shows us that
we must go beyond the SSS regime. In fact the solutions alluded
above are axially symmetric, instead of spherically symmetric,
but seem to
share, with the SSS solutions, the discreetness of the allowed
values of $P_\Delta$\cite{kk}.
Thus we have to go beyond this regime as well.  In fact
there are strong indications (see for example
the discussion in\cite{sud:wal}) that we must go
beyond the static regime, and pose the conjecture in a broad enough
 setting
that would still allow one to single out, for a
given choice of IH charges, a particular black hole
solution and thus a canonical normalization of $\kappa$.
This would be of course the class of stationary black hole solutions,
where we would have to keep track also of the angular momentum,
both at infinity $J_\infty $ and at the horizon $J_\Delta$. The
completeness conjecture would thus be:
$C2$:
{\it For every value of the Isolated Horizon parameters
$a_\Delta, P_\Delta, Q_\Delta, J_\Delta$ for which a space-time
can be constructed, there exist also a  stationary Black Hole Solution
with the same value of the parameters, now characterizing
the Killing Horizon}.

\section{Hair and Solitons}
\label{sec4}
By considering the Hamiltonian formulation for Isolated black holes
in the Static sector, we are
lead  to a formula relating HHM and ADM mass of the colored
BH solutions with the ADM mass of the Solitons of
the theory.

This result is arrived at by the use a general argument from symplectic
geometry that states that, within each connected component of the Static
space embedded in the space of isolated horizons, the value of the
Hamiltonian $H_t$ remains constant\cite{abf1}.
 In particular, it implies that its value is independent of
 the radius $r_\Delta$ of the horizon. Thus, by considering the limit
$ r_\Delta \to 0$, we
and arrive at the following unexpected relation,
\be
M^{(n)}_{\rm ADM}(r_\Delta)
=M^{(n)}_{\rm BK} +
\frac{1}{2}\int_0^{r_\Delta} \,\beta^{(n)}(\tilde{r})\,
\d \tilde{r}\, .
\label{adm2}
\ee
where $ M^{(n)}_{\rm ADM}(r_\Delta) $ is the ADM mass
of the $n$ colored black hole as function of $r_\Delta$,
$M^{(n)}_{\rm BK}$ is the ADM mass
of the $n$ soliton, and $\beta^{(n)}=2r_\Delta\,\kappa^{(n)}$.
The second term at the RHS of Eq.(\ref{adm2}) is the Horizon Mass
$H_\Delta$.

Another point provided by this type of analysis relates to the issue of
the stability: It is only when $M_{\rm ADM}>
M_\Delta$ that the solution can be unstable.
One very clear
example of this is given by the magnetic RN solution,
which can be considered within both the Einstein Maxwell (EM)
theory and the EYM theory. This solution is stable within EM but
unstable within EYM\cite{bizon3}. 
We can understand this surprising fact in terms of the
different values that the Horizon Mass
$M_\Delta$ takes within each theory\cite{acs}.
Let us then suggest a `rule of thumb'
for finding potentially unstable solutions, motivated by the EYM
system. In the static family of solutions, consider the limit
$r_\Delta\mapsto 0$.
We have three possibilities: i) We arrive at a regular
solution with zero energy (i.e. Minkowski). This indicates that
the whole family, labeled by $r_\Delta$, is stable; ii) There
is a minimum allowed value of $r_\Delta$ corresponding to zero
surface gravity. In this case, we can not conclude anything, and;
iii) In the limit one finds a regular solution with positive energy
(a soliton different from the vacuum). In this case,
the whole family of solutions (including
the soliton) is potentially unstable. For a complete discussion
of stability based on energetic considerations see\cite{acs}.

\section{discussion}
\label{sec5}

Let us summarize.  We have studied the extension  of the
Isolated Horizon formalism to include the EYM system and
found that it  provides a powerful
tool for studying some classical aspects of the theory
already at the Static level. In particular,
we found an  unexpected relation between the ADM mass of
a static spherical
black hole solutions, its Horizon mass and the ADM mass of the
corresponding solitonic
solution, and a novel way to  consider the potential instability of
black holes. We have also seen that the IH formalism provides
a framework in which uniqueness (`no-hair') conjectures can be
posed, something that was absent in the standard framework based
in charges at infinity.

\section*{Acknowledgments}
This work was in part supported by
DGAPA-UNAM grant No IN121298, and by CONACyT
grants J32754-E and 32272-E.


\begin{thebibliography}{99}


\bibitem{mh} M. Heusler, \textit{Black Hole Uniqueness Theorems}
(Cambridge University Press: Cambridge, 1996);
 P. Chru\'sciel, 
\textit{Contemporary Mathematics} {\bf 170}, 23 (1994).

\bibitem{prl} A. Ashtekar, C. Beetle, O. Dreyer, S. Fairhurst, B. Krishnan , J. 
Lewandowski, J. Wisniewski, Generic Isolated Horizons and its applications,
\textit{Phys. Rev. Lett.} {\bf 85}, 3564 
(2000).

\bibitem{abf1}  A. Ashtekar, C. Beetle  and S. Fairhurst, 
Isolated Horizons: A generalization of black hole mechanics,  
\textit{Class.\ Quantum Grav.} {\bf 16}, L1-L7 (1999). \\
A. Ashtekar, C. Beetle  and S. Fairhurst, Mechanics of
Isolated Horizons,  \textit{Class.\ Quantum Grav.} {\bf 17}, 253 (2000).\\
A. Ashtekar and A. Corichi, Laws Governing Isolated Horizons: Inclusion of
Dilaton Couplings, \textit{Class. Quantum Grav.} {\bf 17}, 1317 (2000).

\bibitem{ac:ds} A. Corichi and D. Sudarsky, Mass of Colored Black
Holes, \textit{Phys. Rev.} {\bf D61}, 101501 (2000).\\ 
A. Corichi, U. Nucamendi and D. Sudarsky,  Einstein-Yang-Mills
Isolated Horizons: Phase Space, Mechanics, Hair and Conjectures, 
\textit{Phys. Rev.} {\bf D62}, 044046 (2000).



\bibitem{review} M. S. Volkov  and D. V. Gal'tsov D V 
Gravitating Non-Abelian
solitons and black holes with Yang-Mills fields, \textit{Physics Reports}
{\bf 319}, 1-83 (1999). 




\bibitem{kk} B. Kleihaus and J. Kunz, Static black holes with axial
symmetry, \textit{Phys. Rev. Lett.} {\bf 79}, 1595 (1997);

B. Kleihaus and J. Kunz, Static axially-symmetric
Einstein-Yang-Mills-dilaton solutions, 2. black holes
solutions, \textit{Phys. Rev.} {\bf D57} 6138 (1998).


\bibitem{afk} A. Ashtekar, S. Fairhurst and B. Krishnan, Isolated
Horizons: Hamiltonian Evolution and the First Law, \textit{Phys. Rev.}
{\bf D62} 104025 (2000).

\bibitem{sud:wal}  D. Sudarsky and R. Wald,  Extrema of mass,
stationarity, and staticity, and solutions to the Einstein-Yang-Mills 
equations, \textit{Phys.\ Rev.} {\bf D 46}, 1453 (1992).


\bibitem{bizon3} P. Bizon, Stability of Einstein-Yang-Mills black
holes, \textit{Phys.\ Lett} \textbf{B259}, 53 (1991);\\
K. Lee, V.P. Nair and E. Weinberg, A classical instability of
magnetically charged Reissner Nordstorm solutions and the fate of
magnetically charged black holes, \textit{Phys.\ Rev.\ Lett.}
\textbf{68}, 1100 (1992); \\
P.C. Aichelburg and P. Bizon, Magnetically charged black holes and their
stability, \textit{Phys.\ Rev.} {\bf D48}, 607 (1993).
 


\bibitem{acs} A. Ashtekar, A. Corichi and D. Sudarsky, ``Solitons,
Hairy Black Holes and Horizon Mass''. Preprint {\tt gr-qc/0011081}.






\end{thebibliography}
\end{document}